\documentclass[twocolumn,letter]{jpsj2} 

\def\runauthor{Sei \textsc{Suzuki} and Tatsuya \textsc{Nakajima}}

\title{%
Quantum Monte-Carlo method without negative-sign problem\\
for two-dimensional electron systems
under strong magnetic fields
}

\author{%
Sei \textsc{Suzuki}\thanks{E-mail address: sei@sola.c.u-tokyo.ac.jp} and
Tatsuya \textsc{Nakajima}$^{1}$
}

\inst{%
Department of Basic Science, University of Tokyo, Tokyo 153-8902 \\
$^1$Department of Physics, Tohoku University, Sendai 980-8578
}

\recdate{\today}

\abst{%
The quantum Monte-Carlo method is applied to two-dimensional electron
systems under strong magnetic fields. 
The negative-sign problem involved by this method
can be avoided for certain filling factors 
by modifying interaction parameters from those of 
the Coulomb interaction. 
Our techniques for obtaining sign-problem-free parameters 
are described in detail.
Calculated results on static observables are also reported 
 for Landau level filling $\nu = 1/3$.
}

\kword{%
quantum Monte-Carlo, negative-sign problem, linear programming, 
fractional quantum Hall states
}

\begin{document}
\sloppy
\maketitle

It has been revealed that a two-dimensional electron system
in a strong magnetic field has various ground-state phases. 
In particular, the existence of incompressible liquids
gives rise to the fractional quantum Hall 
effect~\cite{bib:FQHE,bib:Pan,bib:book} and
has attracted much theoretical and experimental interests
to this so-called quantum Hall (QH) system.
Rich quantum phases realized in the QH system are attributed 
to Coulomb interactions between electrons, 
because the kinetic energy is quenched 
by the Landau quantization in a strong magnetic field.
Thus the ground state involves strong electronic correlations 
and is difficult to consider by use of mean field theory. 

So far several many-electron states have been proposed theoretically
for the ground state in each phase.
The Laughlin state~\cite{bib:Laughlin}, Jain state~\cite{bib:Jain}, 
and Pfaffian state~\cite{bib:MR} successfully account for
incompressible liquid phases in the QH system.
On the other hand, the composite-fermion liquid 
states~\cite{bib:HLR,bib:RR} and 
anisotropic charge-density-wave states~\cite{bib:Koulakov}
are plausible to compressible phases at the half filling 
of the lowest and higher Landau levels, respectively.

When the properties of such strongly-interacting 
systems are investigated, 
numerical studies can provide a variety of significant information.
The exact diagonalization method of the Hamiltonian matrix 
has been frequently used to inspect the various properties 
of the QH system~\cite{bib:YHL} so far, but 
its applicability is limited to small-size systems.
Recently the density matrix renormalization group (DMRG) method
has been applied to the study of the ground-state properties 
of the QH system~\cite{bib:DMRG}. 
The system size permitted in the DMRG method is much larger than that 
in the exact diagonalization study, 
but it is difficult to study dynamical quantities by the DMRG method.

In the present study, we propose an application of the
quantum Monte-Carlo (QMC) method to the QH system. 
This is because the QMC method can investigate 
not only static but also dynamical quantities in large-size systems.
This method can be applied for both zero- and finite-temperature cases
and has ever been used for a variety of 
quantum many-body systems.
However, the QMC method is accompanied by 
the negative-sign problem, which sometimes makes
us away from investigating physical properties.
Thus we first need to resolve the negative-sign problem
in order to make the best use of the present method.
In this letter, we present one solution 
to the negative-sign problem in the QH system.

This letter is organized as follows. 
We first express physical quantities in the QH system 
in terms of the auxiliary-field path-integral. 
Then we describe how to avoid the negative-sign problem
in the present system. 
Finally some results of QMC calculations are presented.

The system studied here is composed of 
interacting electrons confined on a spherical 
surface~\cite{bib:sphere}. 
It is assumed that a magnetic monopole is located 
at the center of the sphere.
The monopole induces a uniform magnetic field on the surface. 
The number of flux quanta diverging
from the monopole is denoted by $2s$ hereafter ($2s$: integer).
Single-electron states on the sphere are specified 
by the Landau level index and the $z$-component quantum number, $m$,
of angular momentum (AM).
We consider the strong magnetic-field limit and
restrict the single-electron states to the subspace of
the lowest Landau level (LLL).
Then a single-electron state is specified 
by only $m$ which ranges from $-s$ to $s$, and 
the number of degenerate single-electron states 
amounts to $N_{s}=2s+1$. 
The spin degrees of freedom are neglected for simplicity. 

In the LLL approximation, 
the kinetic energy is constant for QH systems 
containing the fixed number of electrons.
The remaining term of the Hamiltonian comes from Coulomb interactions 
between electrons. We express the interaction Hamiltonian
in a bilinear form of the density operator
to introduce the Hubbard-Stratonovich transformation 
needed for QMC calculations.

The density operator is defined using the 
Clebsch-Gordan coefficient as 
$\rho_{K N} = \sum_{m_1=-s}^{s}\sum_{m_2=-s}^{s}
 \langle K N|s m_1; s m_2\rangle a_{m_1}^{\dagger}\tilde{a}_{m_2}$,
where $a_m$ is the annihilation operator of an electron with 
AM quantum number $m$ and $\tilde{a}_m = (-1)^{s+m}a_{-m}$ 
is the time reversal of $a_m$. 
The interaction Hamiltonian is written in terms of the 
density operator as
\begin{equation}
\!\!\!
 \mathcal{H} = -\frac{1}{2}\sum_{K=0}^{2s}\chi_K \sum_{N=-K}^K
  (-1)^{K+N}\rho_{{K N}} \,\rho_{{K -N}} + 
  \epsilon_0 \rho_{{0 0}}.
  \label{eq:Hamiltonian}
\end{equation}
Here $\epsilon_0 = -\sum_{J=0}^{2s}(-1)^{2s-J}(2J+1)
V_J/(2\sqrt{2s+1})$ and 
the coefficient $\chi_K$ is given in terms of 
the Haldane pseudopotential~\cite{bib:sphere1}, $V_J$, as
$\chi_K = \sum_{J=0}^{2s} T_{K J} V_J$. 
The matrix $T$ is explicitly written using Wigner's $6j$ symbol as 
$T_{K J} = \sum_{J=0}^{2s}(-1)^{2s-J+K+1}(2J+1)
\{^{s s J}_{s s K} \}$.

The interaction Hamiltonian can also be written 
in another form as
$\mathcal{H} = \frac{1}{2}\sum_{J=0}^{2s} V_J \sum_{M=-J}^J
 A_{J M}^{\dagger} A_{J M}$, where 
$A_{J M}$ is the pairing operator given by
$ A_{J M} = \sum_{m_1=-s}^s \sum_{m_2=-s}^s 
 \langle J M|s m_1; s m_2\rangle a_{m_1}a_{m_2}$.
Since $a_{m}$ satisfies the 
anti-commutation relation,
$A_{J M}=0$ for even $2s-J$. 
Thus the terms with $V_J$ for even $2s-J$ are unphysical 
and have no influence on physical quantities.
Therefore these unphysical pseudopotentials can be used 
as tunable parameters to improve the efficiency of 
QMC calculations.

The imaginary-time evolution operator
$e^{-\beta\mathcal{H}}$ is decomposed to 
the product of imaginary-time slices
$e^{-\Delta\beta\mathcal{H}}$ ($\beta$: inverse temperature, 
$\Delta\beta \equiv \beta/N_{\rm t}$, 
$N_{\rm t}$: Trotter number).
By introducing the Hubbard-Stratonovich transformation 
for each slice, 
the evolution operator $e^{-\beta\mathcal{H}}$ can be written 
in terms of a linearized Hamiltonian given by
\begin{eqnarray}
 h(\sigma) \!\!\! &=& \!\!\! \frac{1}{2}\sum_{K=1}^{2s}
 \eta_K\chi_K\biggl[
 \sum_{N=1}^K 2 \left( \sigma_{K N}^{\ast}\rho_{K N} + 
 \sigma_{K N}\tilde{\rho}_{K N}
 \right)  \nonumber \\
 &&\!\!\!+\sigma_{K 0}^{\ast}\rho_{K 0} + 
 \sigma_{K 0}\tilde{\rho}_{K 0}\biggr]
 + \frac{1}{2}\chi_0 (\rho_{0 0})^2 + \epsilon_0 \rho_{0 0}. ~~~
 \label{eq:linearh}
\end{eqnarray}
Here $\tilde{\rho}_{K N}=(-1)^K {\rho}_{K N}^{\dagger}$, 
$\sigma_{K N}$ is an auxiliary field for a mode $(K,N)$ 
introduced by the Hubbard-Stratonovich transformation, and
$\eta_K$ is a numerical factor which takes $1$ for non-negative 
$\chi_K$ and $i$ for negative one.
We remark that the terms, $\chi_{0 0} (\rho_{0 0})^2/2$ and 
$\epsilon_0 \rho_{0 0}$,
in Eq.(\ref{eq:linearh}) are constants
because $\rho_{0 0}$ is proportional 
to the fixed number of electrons. 

The partition functions, $Z = {\rm Tr}[e^{-\beta\mathcal{H}}]$
for canonical ensemble and 
$Z = \langle\psi |e^{-\beta\mathcal{H}}|\psi\rangle$ 
for zero temperature,
are represented in terms of auxiliary-field path-integral as
$
 Z = \int \mathcal{D}\sigma \,
 \zeta (\sigma)
$.
Here the weight function is defined by
$\zeta (\sigma) \equiv {\rm Tr} [U(\beta;0)]$
for canonical ensemble and by
$\zeta (\sigma) \equiv \langle\psi | U(\beta;0) |\psi\rangle$
for zero temperature, 
$U(\beta;0)$ in the weight function 
is defined by
$U(\beta;0) \equiv
e^{-\Delta\beta h(\sigma^{\left(N_t\right)})}\cdots
e^{-\Delta\beta h(\sigma^{\left(1\right)})}$,
and $|\psi\rangle$ in the zero-temperature formalism is an 
arbitrary state not orthogonal to the true ground state.
The expectation value of an observable $\mathcal{O}$ 
is written in the path-integral representation as
$
\langle\mathcal{O}\rangle =
\int \mathcal{D}\sigma \langle\mathcal{O} \rangle _{\sigma} 
\zeta (\sigma) / Z
$, where 
$
 \langle\mathcal{O}\rangle _{\sigma} \equiv 
 {\rm Tr}[\mathcal{O} U(\beta; 0)] / \zeta (\sigma)
$ for canonical ensemble and 
$
 \langle\mathcal{O}\rangle _{\sigma} = 
 \langle\psi |U(\beta; \beta /2)\mathcal{O}
 U(\beta /2; 0) |\psi\rangle / \zeta (\sigma)
$ for zero temperature.

The auxiliary-field path-integral is evaluated by means of the
Metropolis-Monte-Carlo technique. 
However, the weight function, $\zeta(\sigma)$, 
can be negative for certain configurations of auxiliary fields. 
Thus let us regard 
$P(\sigma)\equiv |\zeta(\sigma)|$ as 
a positive-definite probability distribution. 
The expectation value 
is then written as
$
 \langle\mathcal{O}\rangle = 
 \int \mathcal{D}\sigma \, [\langle\mathcal{O}\rangle_{\sigma}
 \, \xi(\sigma)]\,P(\sigma)/
 \int \mathcal{D}\sigma \, \xi(\sigma)\,P(\sigma), 
$
where 
$
\xi(\sigma)\equiv \zeta(\sigma)/|\zeta(\sigma)|
$ 
is the sign of $\zeta(\sigma)$.
If the expectation value of $\xi(\sigma)$ is vanishingly small,
Monte-Carlo evaluations of $\langle\mathcal{O}\rangle$ 
become very unstable. 
This is the so-called negative-sign problem.
Thus,
in order to make a precise evaluation, it is desirable that
$\zeta(\sigma)$ is always positive.

For detailed discussion on the negative-sign problem, 
let us move to the matrix representation 
of the QMC method~\cite{bib:QMCBook}.
In order to exploit the time-reversal symmetry,
we first define fermionic operators, $\alpha_m$, by
$\alpha_m = a_m$ for $m\ge 0$, and $\alpha_m = (-1)^{s-m}a_m$ for $m<0$.
By introducing a matrix ${\bf M}(\sigma)$ as 
$ -\Delta\beta \,h(\sigma) \,= \sum_{m,n=-s}^s M_{m n}
(\sigma)\,
 \alpha_m^{\dagger}\alpha_n$,
we define 
${\bf U}(\beta;0) \equiv 
e^{{\bf M}(\sigma^{\left(N_t\right)})}\cdots 
e^{{\bf M}(\sigma^{\left(1\right)})}$.
The weight function in the canonical ensemble
is produced by operating the particle-number projection to that 
in the ground-canonical ensemble~\cite{bib:SMMCPRep}: 
\begin{equation}
 \!\!\! \zeta (\sigma) = \sum_{I=1}^{N_s} 
 e^{- 2\pi i N_e (I/N_s)} 
 \det[{\bf 1}+e^{2\pi i (I/N_s)}{\bf U}(\beta;0)],
 \label{eq:zetac}
\end{equation}
where $N_e$ is the number of electrons.
In the zero-temperature formalism, 
the weight function is given by
$\zeta (\sigma) = \det[{\bf V}^{\dagger}{\bf U}(\beta;0){\bf V}]$, 
where an arbitrary state is written in terms of a matrix ${\bf V}$ 
and electron vacuum $|0\rangle$ as 
$ |\psi\rangle = \left(\sum_{m=-s}^s \alpha_m^{\dagger}V_{m 1}\right)
 \cdots \left(\sum_{m=-s}^s \alpha_m^{\dagger}V_{m N_e}\right)
 |0\rangle$.  
We remark that ${\bf M}$ and ${\bf U}$ are square matrices of dimension 
$N_{s}$, while ${\bf V}$ is a $N_{s} \times N_e$ rectangular matrix.

For the zero-temperature formalism, 
the negative-sign problem
is overcome under the following conditions~\cite{bib:SMMCPRC}:\\
(i) $2s$ is odd and $N_e$ is even.\\
(ii) $\chi_K$ is non-negative for $K=1,2,\cdots ,2s$. \\
By the condition (i), the dimension, $N_{s}$, of matrix ${\bf M}$ 
becomes even.
The condition (ii) yields $\eta_K = 1$ in Eq.(\ref{eq:linearh}) and 
then ${\bf M}$ satisfies
$M_{-m_1,-m_2} = M_{m_1,m_2}^{\ast}$ and 
$M_{-m_1,m_2} = - M_{m_1,-m_2}^{\ast}$ 
for positive $m_1$ and $m_2$. 
Therefore, by arranging AM indices, $m$, 
in the matrix representation as 
$m = s, \cdots, 1/2, -s, \cdots, -1/2$, 
${\bf M}$ has such a form as
\begin{equation}
 {\bf M} = \left[\begin{array}{@{\,}cc@{\,}}
   {\bf A} & {\bf B}  \\
   -{\bf B}^{\ast} & {\bf A}^{\ast}  \end{array}\right],
 \label{eq:matrix}
\end{equation}
where ${\bf A}$ and ${\bf B}$ are square block matrices of dimension
$N_{s}/2$. 
When matrices ${\bf M}_1$ and ${\bf M}_2$ are of the form of Eq.(\ref{eq:matrix}), 
the product ${\bf M}_1 {\bf M}_2$ is also of the form of Eq.(\ref{eq:matrix}).
Thus 
${\bf U}(\beta;0)$ also has the form of Eq.(\ref{eq:matrix}).
When we choose 
a matrix ${\bf V}$ that satisfies 
$V_{-m,N_e/2+n} = V_{m,n}^{\ast}$ and 
$V_{-m,n} = -V_{m,N_e/2+n}^{\ast}$ for positive $m$ and 
$1 \leq n \leq N_e/2$, 
the following property is also obtained:
\begin{equation}
 {\bf V}^{\dagger}{\bf U}(\beta;0){\bf V} =
 \left[\begin{array}{@{\,}cc@{\,}}
   {\bf P} & {\bf Q}  \\
   -{\bf Q}^{\ast} & {\bf P}^{\ast}  \end{array}\right] ,
 \label{eq:VUV}
\end{equation}
where ${\bf P}$ and ${\bf Q}$ are square block matrices of dimension $N_e/2$.
Then, if $\lambda_i$ is an eigenvalue of 
${\bf V}^{\dagger}{\bf U}(\beta;0){\bf V}$, 
its complex-conjugate is also shown to be its eigenvalue. 
Thus we obtain
$\zeta (\sigma) = 
   \prod_{i=1}^{N_e/2} \lambda_i \lambda_i^{\ast}
   \geq 0$.

We can avoid the negative-sign disaster
also in case of the canonical ensemble
by realizing the following conditions: 
(i$'$) $2s$ is odd, and (ii) $\chi_K\geq 0$ for
$K = 1,\cdots,2s$. 
We note that even $N_e$ is not required in (i$'$).
Since the matrix ${\bf M}$ and ${\bf U}(\beta;0)$ then 
have the form of Eq.(\ref{eq:matrix}) as in the zero-temperature 
formalism, the weight function in the ground-canonical
ensemble, $\det[{\bf 1} + {\bf U}(\beta;0)]$, always becomes non-negative.
The weight function in the canonical ensemble, 
which is obtained by operating the particle-number projection 
to that of the ground-canonical ensemble [see Eq.(\ref{eq:zetac})],
is expected to be almost always positive.
We have numerically confirmed that this is the case.


The next task is to inquire which systems satisfy 
the negative-sign-free conditions. 
The condition (i) [(i$'$) for the canonical ensemble] 
restricts the Landau-level filling accessible by the QMC method. 
For instance,
the $\nu = 1/m$ Laughlin state ($m$: odd), characterized by
$2s = m (N_e - 1)$, can be studied by the QMC method if $N_e$ is even.
The condition (ii) brings about strong constraints for electron-electron
interactions accessible by the QMC method.
In fact, this condition is not satisfied by 
the Coulomb interaction within the LLL.

As mentioned above, 
the Haldane pseudopotentials, $V_J$, are classified into the physical
components ($2s-J$: odd) or unphysical ones ($2s-J$: even).
We obtain coupling constants $\chi_K$ favorable to QMC calculations 
by using the unphysical ones as tunable parameters, with
minimizing the difference of the physical ones 
between our model interaction and the Coulomb interaction. 
Then the model pseudopotentials are given in terms of obtained $\chi_K$ 
by $V_J = \sum_{K=0}^{2s} T_{J K} \chi_K$, because
$\sum_K T_{J K} T_{K J^{\prime}} = \delta_{J J^{\prime}}$.
We define the quantity to be minimized by 
$f(-\chi_0,\chi_1,\cdots,\chi_{2s}) = 
 \sum_{2s-J: {\rm odd}} \,\kappa_J 
 \left(
\sum_{K=0}^{2s} T_{J K} \chi_K 
- V_J^{\rm C}\right)$,
where $V_J^{\rm C}$ is the pseudopotential of
the Coulomb interaction and $\kappa_J$ 
controls the priority of each pseudopotential component. 
We note that the arguments of $f$ should be non-negative, 
which agrees with the condition (ii). 
Although the first argument, $-\chi_0$, is not referred by the
condition (ii), 
it must be positive in order to make a repulsive pseudopotential
$V_0 = \sum_{K=0}^{2s} T_{0 K} \chi_K > 0$ 
under the conditions that 
$\chi_K > 0 $ for $1 \le K \le 2s$
and $T_{0 K} < 0$ for all $K$.
Such a set of $\chi_K$ can be obtained by the linear 
programming method, which minimizes $f$ 
over its non-negative arguments imposing the inequalities 
$V_J \geq V_J^{\rm C}$ ($V_J \le V_J^{\rm C}$) 
for non-negative (non-positive) $\kappa_J$.

A simple condition for the linear programming method is
given by $\kappa_J = 1$ for all of odd $2s - J$. 
Then physical pseudopotentials are determined 
with equal weights and become larger than
those of the Coulomb interaction, respectively.
The coupling constants $\chi_K$ obtained in this way vanish
except for the cases of $K=0$ and $K=1$.
The Hamiltonian for this interaction is written as 
$H = - \frac{1}{2}\chi_0 N_e^2/N_s + 6\chi_1 \mathcal{L}^2
/[(N_s-1) N_s (N_s+1)]$, where $\mathcal{L}$ denotes the total 
AM operator.
The eigen-energies of this Hamiltonian depends only on the total AM 
quantum number. Therefore the ground state is energetically degenerate
and the excitation spectrum is gapless.


\begin{figure}[htbp]
\begin{center}
 \includegraphics[width=5cm,clip]{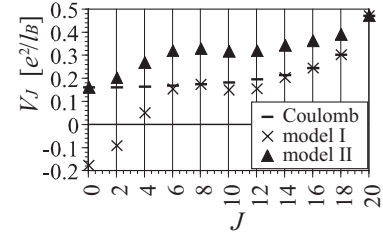}
\end{center}
 \caption{Pseudopotentials for negative-sign-free interactions and
 the Coulomb interaction within the lowest Landau level 
 with $N_{s} = 2s + 1 = 22$. 
 Only physical components for odd $2s - J$ are shown. 
 The pseudopotentials of negative-sign-free model interaction named I 
 are obtained by the linear programming method and 
 are shown by cross symbols. 
 The pseudopotentials of interaction II shown by 
 closed triangles are obtained by applying
 uniform shifts and rescaling to those of interaction I.
}
 \label{fig:interaction}
\end{figure}

Figure \ref{fig:interaction} shows the pseudopotentials of
the Coulomb interaction and two negative-sign-free interactions.
For the interaction I whose pseudopotentials are shown by cross symbols, 
we set the priority parameters as 
$\kappa_{0}=-1$ and $\kappa_{J}=0$ for $J\neq 0$.
Namely we minimized the difference, $|V_0 -V_0^{\rm C}|$, between the 
longest-range components to mimic the long-range nature of 
the Coulomb interaction.
As the inequalities imposed in the linear programming method, 
we used the following ones:
$V_{2s-1} \ge V_{2s-1}^{\rm C}$ (the shortest-range component)
and $V_J \le V_J^{\rm C}$ for $J \ne 2s-1$.
This is because the Laughlin state favors prominently large 
short-range components.
Pseudopotentials of model interaction I 
agree well with those of the Coulomb 
interaction on short-range components, but do not on long-range ones.
In fact, the long-range components 
shift downward from those of the Coulomb interaction 
and eventually become even negative.

In order to make improvements on the interaction I,
we perform the following operations on its pseudopotentials:
(i) uniform shift of pseudopotentials, 
$V_J \to V_J + |V_0^{\rm I}|$, 
(ii) energy rescale, $V_J 
\to V_J \times (V_{2s-1}^{\rm C}-V_0^{\rm C})
/(V_{2s-1}^{\rm I}-V_0^{\rm I})$,
and 
(iii) another uniform shift of pseudopotentials, 
$V_J \to V_J + V_0^{\rm C}$.
As a result of these sequential operations,
we obtain another interaction named II,
whose pseudopotentials are given
by $V_J^{\rm II}=V_0^{\rm C}+(V_{J}^{\rm I}-V_0^{\rm I})
\times (V_{2s-1}^{\rm C}-V_0^{\rm C})
/(V_{2s-1}^{\rm I}-V_0^{\rm I})$.
We note here that uniform shifts of pseudopotentials 
make no influence on the electronic states, because
the change, $V_J \to V_J + v$ for all $J$, 
involves just a constant term in the Hamiltonian as 
$\frac{1}{2}v\sum_{J M}A_{J M}^{\dagger}A_{J M} = 
\frac{1}{2} v N_e(N_e-1)$. 
Furthermore these operations do not break 
the negative-sign-free character of $\chi_K$, 
because uniform shifts of $V_J$ bring about only 
the change in $\chi_{K=0}$
and energy rescale keeps the sign of $\chi_K$.
Thus the interaction II is also free of negative-sign.

The pseudopotentials of interaction II are shown 
by closed triangles in Fig.\ref{fig:interaction}. 
The energy scales of interaction II and Coulomb interaction 
are the same, because $V_{2s - 1} - V_0$ gives the energy scale.
The character of interaction II is similar to that of the Coulomb
interaction on short- and long-range components, 
while the difference between pseudopotentials of the 
interaction II and Coulomb interaction is not so small
on middle-range ones.  
%
In spite of this disagreement on middle-range ones,
it is shown later that 
the interaction II leads to the ground-state properties 
qualitatively similar to those of the Coulomb-interacting system.

\begin{figure}[htbp]
 \begin{center}
  \includegraphics[width=8cm,clip]{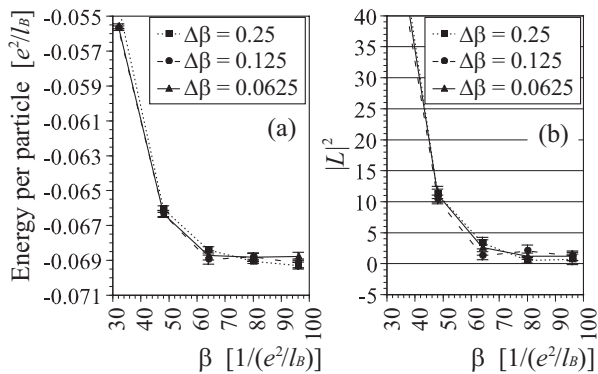}
 \end{center}
 \caption{Expectation values of (a) energy per particle and
 (b) total angular momentum by the zero-temperature QMC calculations
 of the QH system with $2s = 21$ and $N_e = 8$ ($\nu = 1/3$).
The unit of energy is $e^2/l_B$ ($l_B$: the magnetic length),
$\beta$ the inverse temperature, and $\Delta\beta$ is the 
width of imaginary-time slice.
The energy contribution by the positive-charge background 
is also taken into account.
}
 \label{fig:zero}
\end{figure}

Figure \ref{fig:zero} shows 
the results of zero-temperature QMC 
calculations for the QH system with $2s = 21$ and $N_e = 8$ ($\nu = 1/3$).
These calculations were done by using  
the negative-sign-free interaction II in 
Fig.\ref{fig:interaction}.
Calculated results for three imaginary-time slice widths, 
$\Delta\beta=1/4$, $1/8$, and $1/16$ $[1/(e^2/l_B)]$
are almost the same. Hence the numerical errors introduced by the 
Suzuki-Trotter approximation are not so important 
for $\Delta\beta \lesssim 1/4$. 
For the inverse temperature $\beta \gtrsim 60/(e^2/l_B)$, 
both expectation values 
of energy and angular momentum are almost saturated.
The energy gap for the interaction II is found to be 
$\Delta E \simeq 0.09 (e^2/l_B)$ by the exact diagonalization method. 
The inverse temperature needed for convergence is so high
($\beta \gtrsim 6/\Delta E$).
This seems to be due to a high density of states 
for $E \gtrsim \Delta E$.

\begin{figure}[htbp]
 \begin{center}
  \includegraphics[width=8cm,clip]{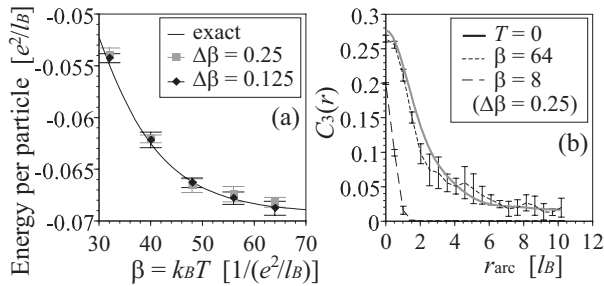}
 \end{center}
 \caption{
Expectation values of (a) energy per particle and
 (b) ODLRO correlation function $C_3(r)$.  Results by the exact 
 diagonalization and  finite-temperature QMC calculations are shown 
 for the QH system  with $2s = 21$ and $N_e = 8$ ($\nu = 1/3$).
$r_{\rm arc}$ denotes the arc-distance 
between ${\textbf r}$ and the north-pole on the Haldane spherical surface.
}
 \label{fig:canonical}
\end{figure}
Figure \ref{fig:canonical}(a)
shows the $\beta$-dependence of energy expectation value
per particle in the canonical ensemble.
The results by the exact diagonalization and QMC calculations 
are in good agreement.
For $\beta \gtrsim 60/(e^2/l_B)$, 
the expectation values of energy per particle 
are almost saturated as well as 
in Fig.\ref{fig:zero}(a). 
Since the existence of the off-diagonal long-range order (ODLRO) 
is one of the most remarkable properties of the Laughlin
state~\cite{bib:ODLRO1},  
we calculated the ODLRO correlation function 
in the $\nu = 1/3$ QH system with interaction II.
The correlation function is defined~\cite{bib:ODLRO2} by
$
C_3({\textbf r}'-{\textbf r}) = \langle A^{\dagger}({\textbf r}')
A({\textbf r})\rangle ,
$
where $A({\textbf r})$ is the annihilation operator of 
a composite boson (an 
electron with three flux quanta attached to it).
Figure \ref{fig:canonical}(b) shows $C_3({\textbf r})$ for 
$\beta = 8$ and $64$ $[1/(e^2/l_B)]$ obtained by QMC calculations
as well as that at $T=0$ obtained by the exact diagonalization.
At zero temperature, $C_3({\bf r})$ remains finite for large separation
$r$. This result suggests that the ground state for the interaction II
should have the ODLRO. For finite temperatures $\beta \lesssim 8$, 
the ODLRO correlation
rapidly vanishes with increasing separation $r$,
while it survives at $\beta = 64 \simeq 6/\Delta E$ 
as in case of $T=0$.
This is because 
the contributions of excited states to $C_3(r)$ is negligible
for temperatures quite lower than $\Delta E$.

We comment on the validity of negative-sign-free interaction II
as a substitute for the Coulomb interaction. 
By the exact diagonalization method, we confirmed that 
the overlap between the two ground states for the interaction II
and Coulomb interaction is not large ($\sim 0.35$) for $\nu = 1/3$.
However, the existence of the ODLRO in the ground state 
and that of energy gap above it 
suggest that an imperfect Bose condensation 
of composite bosons exists for interaction II,
although it is not so perfect as that in the Laughlin state.
Thus we believe that 
the ground state for interaction II
contains the essential properties of the Laughlin state.

The authors thank Y. Kato, A.H. MacDonald, 
and A. Muramatsu for useful discussion.
S.S. acknowledges support by Research Fellowship 
for young scientists of JSPS.
The present work is supported by Grant-in-Aid 
for Scientific Research (Grant No.1406899 and No.14740181) 
by the Ministry of Education, Culture, Sports, Science and 
Technology of Japan.


\end{document}